\input epsf.tex

%
%
\hyphenation{Pij-pers}
%
\newcount\eqnumber
\eqnumber=1
\def\neqn{{\rm(\the\eqnumber)}\global\advance\eqnumber by 1}
\def\refeq#1){\advance\eqnumber by -#1 {\rm(\the\eqnumber)} \advance
\eqnumber by #1}
\def\eqnam#1#2{\immediate\write1{\xdef\ #2{(\the\eqnumber}}\xdef#1{(\the\eqnumber}}
\newcount\fignumber
\fignumber=1
\def\nfig{\global\advance\fignumber by 1}
\def\refig#1{\advance\fignumber by -#1 \the\fignumber \advance\fignumber by #1}
\def\fignam#1#2{\immediate\write1{\xdef\ #2{\the\fignumber}}\xdef#1{\the\fignumber}}
\def\note #1]{{\bf #1]}}
\def\draft{\headline{\bf File: \jobname\hfill DRAFT\hfill\today}}
\def\ref{\par\noindent
	\hangindent=0.7 truecm
	\hangafter=1}
%
%

\draft

\MAINTITLE{Research Note : The residence time of dust grains in turbulent
molecular clouds}


\AUTHOR={ F.P. Pijpers }

\OFFPRINTS{F.P. Pijpers}

\INSTITUTE{
Theoretical Astrophysics Center, Institute for Physics and Astronomy, 
Aarhus University, Ny Munkegade, 8000~Aarhus~C, Denmark }

\DATE{Received ; accepted}
\ABSTRACT{ The residence time of dust grains inside turbulent molecular 
clouds is calculated . Earlier estimates of
this time by Boland \& de Jong (1982) have been criticized by
Prasad et al. (1987) for not taking into account properly the turbulent
character of the velocity field in molecular clouds which implies
that the trajectory of a dust grain should be treated as a random walk.
Taking some minimal assumptions regarding the turbulent velocity field
a time scale is derived that depends on the Reynolds number of the flow.
The near proportionality with Reynolds number of this time scale results 
in a much longer time scale over which dust grains will remain inside a 
molecular cloud.}


\THESAURUS{08(09.04.1 - 09.03.1 - 02.20.1)}

\maketitle

\titlea{Introduction}

Chemical surface reactions on dust grains form an essential pathway in
the formation of complex molecules, and even some quite simple ones, 
in interstellar molecular clouds.
The time that a grain resides inside clouds with a high optical depth
determines its life time and therefore the time over which it can play
a role in the chemical processing of its host molecular cloud. 
This residence time scale has been estimated previously by Boland \& 
de Jong (1982) to be $L/\sigma_v$ from the observables $L$ which is the
size of the cloud and $\sigma_v$ which is its velocity dispersion.
Prasad et al. (1987) argued that this essentially implies a linear path
through the cloud for any dust grain. This does not properly account for
the fact that the velocity field in the cloud is turbulent and that the 
grains, dragged along by the gas motions, perform a random walk. 
Using a constant step size random walk Prasad et al. (1987) derive a time scale
$L^2 /\sigma_v l_s$ where $l_s$ is the (typical) step size in the random walk,
which they estimate to be the scale of the largest eddies in the cloud.
However using a single scale eddy size to represent the large range of 
eddy sizes present in highly turbulent flow is also inappropriate.
Taking dust grains to be test particles of the turbulent flow, it is
possible to derive a circulation time scale in terms of the Reynolds
number of the flow.
From the study of Weidenschilling and Ruzmaikina (1994) it is clear that 
in particular sonic turbulence strongly affects grain growth and destruction 
since it affects the rate of collisions and the energy of such collisions 
with other grains and with gas phase material. The point of focus of this 
paper is not the growth and destruction of grains but rather the stochastic 
migration of an ensemble of grains throughout a dense cloud. It is 
demonstrated that for quite general properties of the turbulence the 
random walk through a cloud indeed considerably increases the time grains 
stay in a region of high optical depth, in accordance with the argument of
Prasad et al. (1987) although the time scale derived here is different 
from theirs.

\titlea{Power law turbulence}
 
The coupling between gas and dust grains in stellar winds and in the 
interstellar medium can be shown to be quite strong, because of the 
strong drag that dust grains experience when drifting with respect to 
the gas. The dust therefore is treated here as completely momentum coupled 
to the gas : effectively it is a test particle for the turbulent flow.
 
The length scale of dissipation (internal scale) of the turbulence $l_d$ 
is set by the viscosity in the gas of the molecular cloud or the magnetic 
diffusivity (the inverse of the conductivity) if the cloud contains a 
(tangled) magnetic field. Unless this length scale of dissipation of \
the turbulent velocity field is nearly equal to the size of the molecular 
cloud the energy density $E$ in the velocity field per length scale 
interval ${\rm d} l$, or equivalently per wave vector interval 
${\rm d} k$, will have an `inertial range' over which it behaves 
as a power law as a function of length scale. Outside of this range the 
energy density or power in the turbulent velocity field decreases rapidly, 
either due to dissipation or because the cloud is not large enough to 
contain the length scales larger than its own size. For stationary 
hydrodynamical turbulence in the absence of magnetic fields, and in the 
inertial range, the energy density has a slope of this power law which 
conforms to the well known Kolmogorov-Obukhov scaling~: $E(k) {\rm d}k 
\propto k^{-5/3} {\rm d}k$ (cf. Landau \& Lifshitz, 1987).
As was pointed out by Scalo (1987) molecular clouds are highly magnetized 
and subject to self-gravity. Also energy is injected into the turbulent 
velocity field from a variety of sources on a variety of spatial scales, 
for instance shear due to galactic rotation, so the slope of the energy
spectrum can differ from the value of ${-5/3}$. 
However even for magnetohydrodynamic (MHD) turbulence the velocity field 
behaves as a power law over a large range of length scales (cf. 
Stani\v si\'c, 1985). Gravity
does not impose a characteristic length scale nor do effects of 
compressibility. The energy injection mechanisms can be subject to one or
more characteristic length scales which could introduce humps or roll-overs
in the turbulent spectrum, where locally the energy spectrum is not a power 
law. However the requirement of stationarity of the turbulent spectrum imposes 
strong constraints on the transfer of energy between scales. Only if the 
turbulent cascading of energy is the least effective (slowest) of the 
energy transfer mechanisms operating, which is unlikely, will the spectrum 
be severely affected by the other effects, because then energy can build
up at the injection length scale before the energy cascading from larger 
scales and to smaller scales can compensate for this. Summarizing, the 
energy density probably does not follow exactly a power law with a slope 
equal to $-5/3$ over the entire range of scales possible, but the 
departures from this behaviour are not likely to be large.

Observationally it is relatively straightforward to determine a power 
law spectrum for the mass $M$ contained in clumps within molecular clouds, 
so the starting point is this mass spectrum, where $f(M)$ is the fraction 
of cloud clumps with a mass between $M$ and $M+{\rm d} M$~:
$$
f(M) {\rm d} M \propto M^{-\alpha} {\rm d} M
\eqno\neqn
$$
Recent measurements of $\alpha$ by Williams et al. (1995) yield a
value for $\alpha = 1.27 \pm 0.09$ for the Rosette Nebula.
Values of up to $1.7$ for this and other cloud complexes have also been 
reported (cf. Blitz, 1991). It appears that $\alpha$ is roughly constant 
over a range of scales from $100\ {\rm pc}$ down to $10^{-4}\ {\rm pc}$
in at least one cloud (cf. Rouan et al. 1997).

\titleb{In the absence of density-velocity correlation}

The mass is related to the gas density $\rho$ and $l$ by~:
$$
M \propto \rho l^3
\eqno\neqn$$
In principle the gas density could be a function of the length scale
$l$, which is discussed in Sect. 2.2. Assuming a $\rho$ 
independent of $l$ the mass spectrum can be converted to a spectrum
in terms of sizes between $l$ and $l+ {\rm d} l$~:
\eqnam\distrib{distrib}
$$
f(l) {\rm d} l = C_L l^{2-3\alpha} {\rm d} l
\eqno\neqn$$
Outside of this range the amount of material is assumed to be negligible 
and the value of the power-law index $\alpha$ is assumed constant over 
the entire inertial range. Even though the observed clumps should probably 
not be individually identified with turbulent eddies it is likely that 
the turbulent motions follow a similar distribution of energy density 
versus size, since the clumps are a consequence of random convergence of
the turbulent flow. In this paper it is assumed that the distributions 
are identical. The constant $C_L$ is determined by demanding that
the integral of $f(l)$ over $l$ between the dissipation length scale $l_d$
and the size of the molecular cloud $L$ be equal to unity. 
$$
\int\limits_{l_d}^{L} C_L l^{2-3\alpha} {\rm d} l = 1
\eqno\neqn$$
implies~:
$$
\eqalign{
C_L\ &=\ \left(3\alpha -3\right) {L^{3\alpha -3} \over R_e^{9\alpha -9
\over 4} -1} \hskip 1cm \alpha\neq 1 \cr
C_L\ &=\ {4 \over 3\ln R_e} \hskip 3cm \alpha = 1 \cr}
\eqno\neqn$$
where the Reynolds number $R_e$ is related to $L$ and $l_d$ by~:
\eqnam\Reydef{Reydef}
$$
R_e\ =\ \left({L\over l_d}\right)^{4/3}
\eqno\neqn
$$
If a cloud is optically thin in the spectral line in which the velocity 
dispersion is measured the observed velocity dispersion of a cloud must be the
expectation value of the square of the velocity. If the cloud is optically
thick the line only measures the turbulence in the surface layers.
Since the turbulent velocity field is probably not homogeneous a velocity
dispersion determined with an optically thick line is then not representative
of the interior of the cloud. In the following it will be assumed that
a measured velocity dispersion is representative however, so that either
the method with which it is measured is sensitive to velocity fields 
throughout the clouds or the turbulence is assumed to be homogeneous
and isotropic. In the usual turbulent cascade (cf. Landau \& Lifshitz, 
1987) of fully developed free turbulence the velocity $v_l$ of a turbulent 
eddy with size $l$ is~:
\eqnam\freetur{freetur}
$$
v_l = v_L \left( {l\over L }\right)^{1/3}
\eqno\neqn$$
where $v_L$ is the overturning velocity of the largest eddies supported by
the molecular cloud which have a size of the order of the cloud size $L$.
As argued at the start of this section a number of effects could change
the power $1/3$ in relation \freetur), but the departures are not likely
to be very large. The velocity dispersion $\sigma_v^2$ is the expectation 
value of the square of the turbulent velocity~:
\eqnam\VelDis{VelDis}
$$
\eqalign{
\sigma_v^2\ &=\ \int\limits_{l_d}^{L} C_L v_l^2 l^{2-3\alpha} {\rm d} l \cr
   &=\ C_L v_L^2 L^{3-3\alpha} {3\over 11-9\alpha}\left(1 - R_e^{9\alpha-11
  \over 4}\right) \hskip 6mm \alpha\neq {11\over 9}\cr
   &\approx \left\vert {9\alpha- 9 \over 9\alpha - 11}\right\vert v_L^2 R_e^{\gamma}
  \hskip 3.3cm \alpha\neq 1, {11\over 9}\cr}
\eqno\neqn$$
In the approximate equality it has been assumed that the Reynolds number
$R_e \gg 1$. The exponent $\gamma$ satisfies~:
$$
\gamma = \cases{
-1/2 \hskip 1.6cm \alpha > {11\over 9} \cr
{9 -9\alpha\over 4} \hskip 1cm 1< \alpha < {11\over 9} \cr
\hskip 3mm 0 \hskip 1.9cm \alpha < 1 \cr
}
\eqno\neqn$$
Note that the intrinsic thermal width of the line is assumed
to be much smaller than the width of the velocity distribution, which 
implies that the turbulent velocities are supersonic. If
this is not the case a proper convolution should be done of the line 
profile and the velocity distribution. 

As a grain is swept along in the turbulent eddies it performs a random
walk through the cloud. If all the steps were of equal length $l$ then
it would take $N_l = (L/l)^2$ steps to traverse a distance $L$, each
step taking a time $T_l = l/v_l$. The step size obeys the
distribution \distrib) and so the total residence time for a dust grain
must be the expectation value of $N_l \times T_l$~:
\eqnam\TimeEst{TimeEst}
$$
\eqalign{
T_{tot}\ &=\ \int\limits_{l_d}^{L} C_L N_l T_l l^{2-3\alpha} {\rm d} l \cr
   &= C_L {L^{4-3\alpha} \over v_L}{3\over\vert 5 - 9\alpha\vert}\left(
  1-R_e^{9\alpha-5\over 4}\right) \hskip 1cm \alpha\neq {5\over 9}\cr
  &\approx \left\vert {\left(9\alpha - 9\right)^3\over\left(9\alpha -11\right)
  \left(9\alpha -5\right)^2}\right\vert^{1/2} {L\over \sigma_v}R_e^{\delta}
  \equiv C_T {L\over \sigma_v}R_e^{\delta}\cr
  &\hskip 4cm \alpha\neq {5\over 9}, 1, {11\over 9}\cr}
\eqno\neqn$$
where the exponent $\delta$ satisfies~:
$$
\delta = \cases{
\hskip 4mm 3/4 \hskip 2.4cm \alpha > {11\over 9} \cr
(17 -9\alpha)/ 8 \hskip 1cm 1< \alpha < {11\over 9} \cr
(9\alpha -5)/ 4 \hskip 1.2cm {5\over 9} < \alpha < 1 \cr
\hskip 6mm 0 \hskip 2.6cm \alpha < {5\over 9} \cr
}
\eqno\neqn$$
The slope $\alpha$ of the mass spectrum of the clouds appears in
the exponent $\delta$ only for ${5\over 9}<\alpha < {11\over 9}$. For 
$\alpha = 1.27$ the exponent $\delta = 3/4$ and $C_T \approx 0.9$. 
For steeper ($\alpha > 1.27$) spectra $C_T$ approaches unity asymptotically. 
It is only for mass spectra that are flat or rising ($\alpha < 0$) that 
the estimate of Boland \& de Jong (1982) is appropriate, because only then
is $\delta = 0$.

Of interest is the dependence of the residence time $T_{tot}$ on the Reynolds
number $R_e$ for $\alpha > 5/9$. It seems perhaps counter-intuitive 
that for more vigorous turbulence (larger $R_e$) dust grains take longer 
to traverse a cloud. However one should realize that if the turbulence 
is more vigorous the velocity dispersion $\sigma_v$ also increases. For 
a given (measured) velocity dispersion an increased Reynolds number 
implies that $l_d$ is smaller so that there is a larger contribution to
the integrals \VelDis) and \TimeEst) from small scales. Thus a grain is 
likely to make many more small steps in a random walk than large ones.

\titleb{Fully coupled density-velocity fluctuations}

In Sect. 2.1 the density $\rho$ is assumed to be constant, so that 
it can be ignored in the integral for the velocity dispersion.
In view of the Larson relations (Larson, 1981) which show that the 
density of clumps follows a power law as a function of clump size,
it is instructive to consider also the effect of a gas density which 
fluctuates in a manner that is spatially correlated with the turbulent
velocity and which has a power law spectrum of $l$~: $\rho\propto 
l^{-\beta}$. The measured value of $\beta = 1.3 \pm 0.2$ (cf. Cernicharo, 
1991 and references therein). The combined distribution function is 
assumed to be the product of the individual distribution functions
for the density and the velocity, which implies that \distrib) becomes~:
$$
f(l) {\rm d} l \propto l^{2-3\alpha+\beta(\alpha -1)} {\rm d} l
\eqno\neqn$$  
In this paper the clouds are not assumed to be in virial equilibrium, 
thus no a-priori relation between $\rho_l$ and $v_l$ is postulated,
and $\beta$ is a free parameter. 
For a cloud that is in virial equilibrium on every scale one expects 
that the kinetic and potential energies of turbulent eddies are
identical up to a constant factor. This implies that $v_l^2 \propto 
M_l/l \propto \rho l^2 \propto l^{2-\beta}$. Combining this with \freetur) 
shows that then one would expect $\beta = 4/3$, which is equal to the 
value actually observed to within $1\sigma$. 

In the analysis presented here $\sigma_v$ depends on $\rho$ since the 
density plays a role in the determining the profile of the spectral 
line from which the velocity dispersion is measured. 
If one takes a density weighted average instead of \VelDis) then~:
\eqnam\VelDisW{VelDisW}
$$
\eqalign{
\sigma_v^2\ &=\ {1\over \langle\rho \rangle}\int\limits_{l_d}^L C_L 
\rho_l v_l^2 l^{2-3\alpha+\beta (\alpha -1)}{\rm d} l \cr
&=\ v_L^2 {\rho_L \over \langle\rho \rangle} C_L L^{3-3\alpha +\beta 
(\alpha -1)} { 1 - R_e^{\left(9\alpha - 9 - 3 \beta (\alpha -2)\right)/4}\over
11 - 9\alpha + 3\beta (\alpha -2)}\cr
&\hskip 2.5cm \alpha\neq{6\beta -11\over 3\beta -9} \cr}
\eqno\neqn$$
where $\langle\rho\rangle$ is the expectation value for the density.
$$
\langle\rho \rangle\ =\ \int\limits_{l_d}^L C_L \rho_l l^{2-3\alpha+
\beta (\alpha -1)}{\rm d} l 
\eqno\neqn$$

\fignam{\powplo}{powplo}
\begfig 5.8cm
\vskip -5.8cm\hskip 6mm
\epsfxsize=6.5cm
\epsfbox{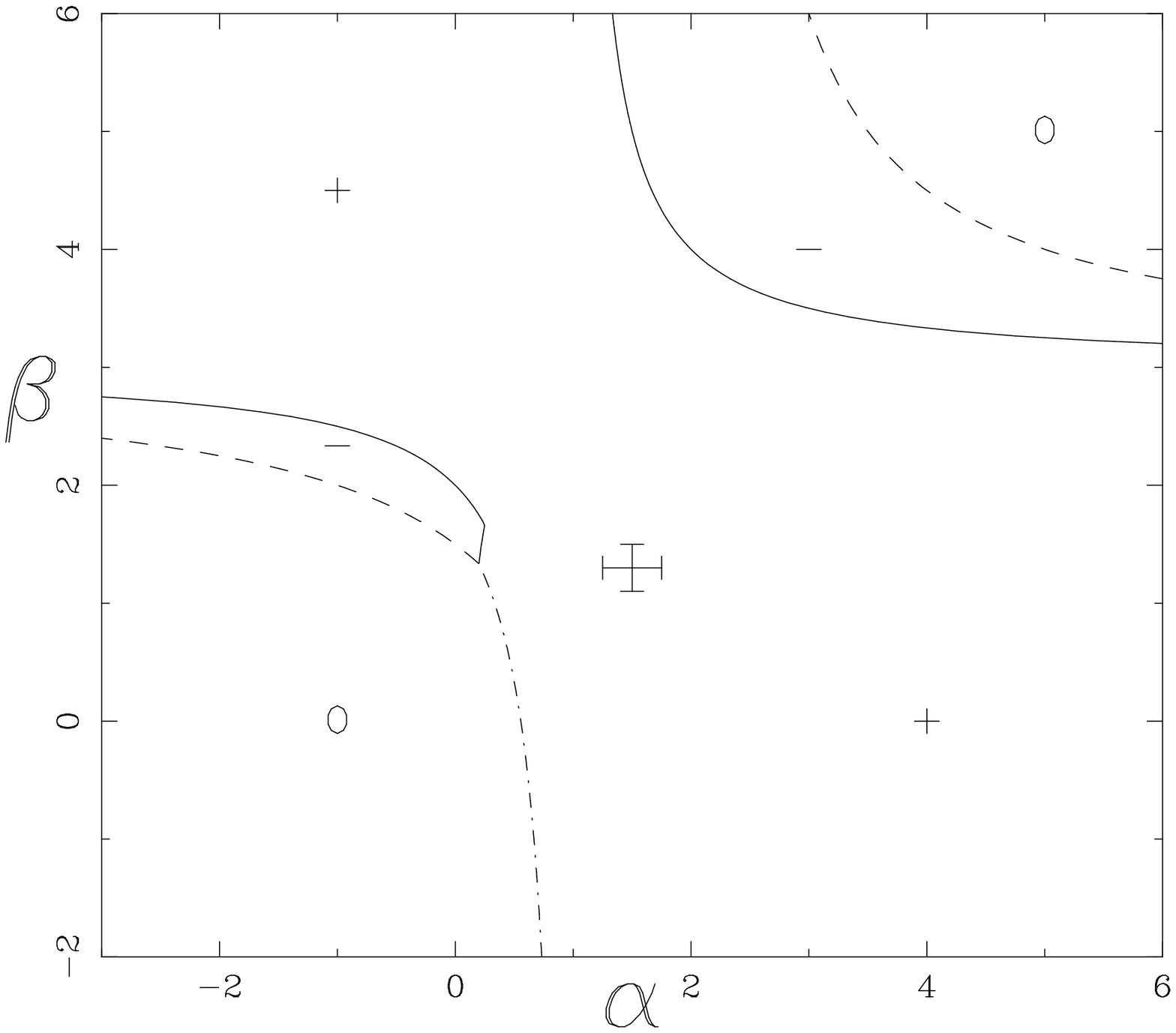}
\figure{\powplo}{$T_{tot}$ is a power law of $R_e$ for $R_e \gg 1$. The
sign of the power law index is shown as a function of the power-law 
exponents $\alpha$ and $\beta$. It is positive in areas indicated by a 
$+$, negative where there is a $-$ and zero on the solid curves and in 
areas indicated by $0$. On the dashed curves $T_{tot}\propto\left(\ln 
R_e\right)^{-1/2}$. On the dash-dotted curve $T_{tot}\propto\ln R_e$.
The error bars show the range of measured values for $\alpha$ and $\beta$ 
for clumps in molecular clouds }
\endfig
\nfig

Note that there is a problem with the velocity dispersion that results
from this analysis. Observations appear to indicate that $\sigma_v 
\propto L^{0.3}$ (cf. Scalo, 1987), whereas with the observed values of 
$\alpha$ and $\beta$ one obtains $\sigma_v \propto v_L^2 R_e^{-1/2}$ which is 
independent of $L$ because of relations \Reydef) and \freetur).
One way to explain this is that when observing at increasing resolution 
(smaller $L$) one concentrates on `clumps', which are the regions with 
higher density and lower velocity dispersion. This `selection effect' 
might introduce the observed trend, bringing it closer to relation \freetur).

The relation \TimeEst) becomes now~:
$$
\eqalign{
T_{tot} &=\ \int\limits_{l_d}^{L} C_L N_l T_l l^{2-3\alpha} {\rm d} l \cr
&= {L\over \sigma_v} {\sigma_v\over v_L} C_L L^{3-3\alpha +\beta
(\alpha -1)} { 1 - R_e^{\left(9\alpha - 5 - 3 \beta (\alpha -1)\right)/4}\over
5 - 9\alpha + 3\beta (\alpha -1)}\cr
&\hskip 2.5cm \alpha\neq{3\beta -5\over 3\beta -9} \cr}
\eqno\neqn$$
It is not particularly illuminating to show the equations for all possible
values of the exponents $\alpha$ and $\beta$, since this separates out 
into 73 separate cases. Instead in Fig. \powplo{} is shown the behaviour 
of the exponent of the Reynolds number $R_e$ as a function of $\alpha$ 
and $\beta$. Indicated by $+$, $0$, and $-$ are regions where this exponent 
is positive, zero, or negative respectively. For $2/3 <\beta <3$ and 
$\alpha >1$ and for $\beta <2/3$ and $\alpha >2$ the dependence of $T_{tot}$
on the Reynolds number is always $T_{tot} \propto R_e^{3/4}$ for $R_e\gg 1$. 
From Fig. \powplo{} it is clear that only for quite steep distributions for
the density (large $\beta$) the exponent is negative and $T_{tot}$ decreases
as $R_e$ increases. In such cases the velocity dispersion is weighted 
strongly towards the high density, low velocity clumps so that for more 
vigorous turbulence the estimate $T_{tot} = L/\sigma_v$ is increasingly 
an overestimate of the residence time. For the measured $\alpha=1.27$, 
$T_{tot}$ is a decreasing function of $R_e$ only for $\beta > 6.7$. 
For the observed range of values $\alpha$ and $\beta$ the relation is
$$
T_{tot} = C_T {L\over \sigma_v} R_e^{3/4} = C_T {L^2 \over \l_d \sigma_v}
\eqno\neqn$$
where the value of $C_T \approx 0.35$.
This appears to be very similar to the expression of Prasad et al. (1987)
who have $l_s$ replacing $l_d$, where $l_s$ is the typical eddy size.
The dissipation length scale $l_d$ is much smaller than the typical
eddy size $l_d \ll l_s$ however, so that the time scale derived here is 
longer than that derived by Prasad et al. (1987). 

\titleb{The influence of fractal structures}

It is important to note that the residence time in the cloud complex 
derived here may be an over-estimate of the residence time of the grains. 
There is growing evidence that the gas density distribution in molecular 
cloud complexes is highly inhomogeneous or `fractal' in structure 
(cf. Scalo, 1987, 1990; Falgarone et~al., 1991). Even though the grain 
can take a long time to traverse a cloud complex it will encounter 
regions within that complex that may well be sufficiently warm or exposed 
to ultraviolet (UV) radiation to be quite destructive for grains or at 
least their icy mantles.
This affects the statistics of the distributions and the various expectation 
values evaluated in this paper. For instance the velocity dispersion (Eq.
\VelDis){}) is independent of the external scale and the density weighted 
velocity dispersion \VelDisW) is as well for any value $\beta \ge 3$, 
whereas it generally will not be for a fractal molecular cloud.
In this case one should properly calculate the characteristic time it 
takes for grains to circulate into exposed regions from dense cold parts
of the cloud core where the grain is shielded. The time of survival 
should then depend on the fractal dimensions of the hospitable and 
inhospitable parts of the cloud. One could
think of this as considerably reducing the Reynolds number because in
definition \Reydef), the value for $L$ should be the size of the high
density (hospitable) regions. This means that for a fractal cloud $L$ 
and the Reynolds number itself become stochastic quantities, for which 
expectation values must be calculated. If the regions with sufficiently 
high density occupy relatively small regions of space, e.g. if these 
regions are filamentary in structure, this reduces $L$ on average
and therefore also $R_e$ and thus $T_{tot}$. However the Reynolds number 
cannot be as low as to be of order unity on average over a molecular 
cloud because that would imply that on average the flow in a cloud complex 
is laminar which is clearly not the case. Therefore the estimate of 
Boland \& de Jong (1982) of $T_{tot} = L/\sigma_v$ for the residence 
time of grains in molecular clouds is too low.

\titlea{Conclusions}

It is shown in this paper that $t = L/\sigma_v$ severely under-estimates 
the actual time spent by grains in turbulent molecular clouds. In 
the limit of large Reynolds number for the turbulence, in clouds in which
the velocity dispersion deduced from line widths represents well the 
velocity distribution throughout the cloud, the residence time is 
$T_{tot} \approx 0.35 R_e^{3/4} L / \sigma_v$. Unfortunately the Reynolds number
is not a directly observable quantity. Using the definition \Reydef) one
can make a crude estimate however. The dissipation scale $l_d$ is determined
by the microphysics of the mechanism responsible for the dissipation of
the turbulence such as viscosity or magnetic diffusivity. Such mechanisms 
usually become effective on length scales comparable to the mean free path of gas 
atoms or molecules. On larger scales the turbulence is essentially free
and nearly dissipationless.
Even at the low densities and low ionization fractions of cold interstellar 
matter this mean free path is not likely to be larger than $0.1\ {\rm pc}$,
at which scale ambipolar diffusion starts affecting the propagation of
hydromagnetic waves (Mouschovias, 1991). 
The size $L$ of a typical molecular cloud is more than $10\ {\rm pc}$. 
Realistic values of the Reynolds number for molecular clouds can thus 
be in excess of $10^3$, and have even been estimated to be of the order 
of $10^6$ (Scalo, 1987), which implies that the residence time of grains 
can be more than $10^2$ or even $10^4$ times longer than previously estimated. 

Taking as an example the clumps in the Rosetta molecular cloud which
have a velocity dispersion of typically $\sim 1.5 {\rm\ km\ /s}$ and a 
size of typically $\sim 1.5 {\rm\ pc}$ (cf. Williams et al., 1995) the 
time scale, for a Reynolds number of $10^4$, becomes $\sim 3\ 10^8 {\rm\ yr}$.
This is already of the order of the life time of a cloud, so grains formed
in the centre of clouds like this will barely manage to migrate out of
the cloud, let alone out of clouds that have a larger Reynolds number.
One might conclude that therefore molecules caught or formed on grains 
will not be returned to the ISM at all, since grains will not reach regions
where photodesorption can play a role. However other effects might intervene.
Vigorous turbulence is likely to lead to a higher frequency of grain-grain 
collisions, and also leads to heating because of dissipation of the 
turbulent energy. Inhomogeneity of a cloud produces a higher UV 
radiation field within the cloud than the cloud would have if it were 
homogeneous. What should be clear from this paper is that all physical 
and chemical processes in which grains are involved may well be modified by
the presence of turbulence, and most certainly will have a much longer time
to operate than estimated previously.

\acknow{The Theoretical Astrophysics Center is a collaboration 
between Copenhagen University and Aarhus University and is funded by 
Danmarks Grundforskningsfonden. Dr. David Field is thanked for helpful
comments on the manuscript.}

\begref{References}

\ref
Blitz L., 1991, 
in `The physics of star formation and early stellar evolution', 
eds. C.J. Lada, N.D. Kylafis,
Kluwer, Dordrecht, 3
\ref 
Boland W., de Jong T., 1982, 
{ApJ}
261, 110
\ref
Cernicharo J., 1991, 
in `The physics of star formation and early stellar evolution', 
eds. C.J. Lada, N.D. Kylafis,
Kluwer, Dordrecht, 287
\ref
Falgarone E., Phillips T.G., Walker C.K., 1991,
{ApJ}
378, 186
\ref 
Landau L., Lifshitz E.M., 1987, 
Course in theoretical physics 6 : Fluid Mechanics, $2^{nd}$ Ed., 
Pergamon, Oxford, 133, 134
\ref
Larson R.B., 1981,
{MNRAS}
194, 809
\ref
Mouschovias T.Ch., 1991, 
in `The physics of star formation and early stellar evolution', 
eds. C.J. Lada, N.D. Kylafis,
Kluwer, Dordrecht, 449
\ref 
Prasad S.S., Tarafdar S.P., Villere K.R., Huntress W.T., 1987,
in `Interstellar Processes', 
eds. D.J. Hollenbach, H.A. Thronson,
Kluwer, Dordrecht, 631
\ref
Rouan D., Field D., Lemaire J.-L., et al., 1997,
{MNRAS}
284, 395
\ref
Scalo J.M., 1987, 
in `Interstellar Processes', 
eds. D.J. Hollenbach, H.A. Thronson,
Kluwer, Dordrecht, 349
\ref
Scalo J.M., 1990, 
in `Physical processes in fragmentation and star formation', 
eds. R. Capuzzo-Dolzetta, C. Chiosi, A. Di Fazio, 
Kluwer, Dordrecht, 117
\ref
Stani\v si\'c M.M., 1985, 
The mathematical theory of turbulence, 
Springer, Berlin, 301
\ref
Weidenschilling S.L., Ruzmaikina T.V., 1994
{ApJ}
430, 713
\ref 
Williams J.P, Blitz L., Stark A.A., 1995,
{ApJ}
451, 252
\endref

\bye